\documentclass[twocolumn,showpacs,preprintnumbers,amsmath,amssymb,superscriptaddress]{revtex4-1}

\usepackage[english]{babel}
\RequirePackage[T1]{fontenc}
\RequirePackage{times} 

\usepackage{amsfonts}
\usepackage{subfigure}
\usepackage{amsmath}
\usepackage{amssymb}
\usepackage{graphicx,epstopdf}
\usepackage{array}
\usepackage{color}
\usepackage{my_symbols}
\usepackage{CJK}
\usepackage{bm}
\usepackage[normalem]{ulem}
\usepackage{hyperref}
\usepackage{color}
\usepackage{mathrsfs}

\hypersetup{
	colorlinks=true,
	linkcolor=blue,
	filecolor=magenta,
	urlcolor=cyan,
}

{\par}

\newcommand\rmv{\bgroup\markoverwith {\textcolor{red}{\rule[0.5ex]{2pt}{0.4pt}}}\ULon}

\begin{document}

\title{Topological Phases in Magnonics}

\author{Fengjun Zhuo}
\email[~]{zhuofj@shanghaitech.edu.cn}
\affiliation{School of Physical Science and Technology, ShanghaiTech University, Shanghai 201210, China}

\author{Jian Kang}
\affiliation{School of Physical Science and Technology, ShanghaiTech University, Shanghai 201210, China}

\author{Aur\'{e}lien Manchon}
\affiliation{Aix Marseille Univ, CNRS, CINAM, Marseille 13288, France}

\author{Zhenxiang Cheng}
\email[~]{cheng@uow.edu.au}
\affiliation{Institute for Superconducting and Electronic Materials, Australian Institute of Innovative Materials, University of Wollongong, Innovation Campus, Squires Way, North Wollongong, NSW 2500, Australia}

\begin{abstract}
Magnonics or magnon spintronics is an emerging field focusing on generating, detecting, and manipulating magnons. As charge-neutral quasi-particles, magnons are promising information carriers because of their low energy dissipation and long coherence length. In the past decade, topological phases in magnonics have attracted intensive attention due to their fundamental importance in condensed-matter physics and potential applications of spintronic devices. In this review, we mainly focus on recent progress in topological magnonics, such as the Hall effect of magnons, magnon Chern insulators, topological magnon semimetals, etc. In addition, the evidence supporting topological phases in magnonics and candidate materials are also discussed and summarized. The aim of this review is to provide readers with a comprehensive and systematic understanding of the recent developments in topological magnonics.  

\end{abstract}

\maketitle

\section{Introduction: Topology Meets Magnon}
Since the discovery of the Giant Magneto-Resistance (GMR) effect in magnetic metallic multilayers \cite{Baibich1988,Binasch1989}, extensive studies on magnetic materials, in particular, ferromagnets and antiferromagnets, have yielded various interesting and remarkable results that form the basis for a new scientific field called spintronics  in recent decades \cite{Wolf2001,Zutic2004,Jungwirth2016,Baltz2018}. Spintronics explores the coupled electron spin and charge transport in magnetic materials, and has attracted intensive attention for its fundamental interest and potential impacts in logic operations and data storage devices \cite{Atkinson2003,Allwood2005,Parkin2008}. Compared with the traditional electronic devices in semiconductors, spintronics has the advantages of nonvolatility, ultrafast data processing speed, ultrahigh data storage density, and less electric power consumption \cite{Wolf2001}. Nowadays diverse new concepts in spintronics have sprung up, such as spin transfer torques \cite{Slonczewski1996,Berger1996}, spin Hall effect \cite{Dyakonov1971a,Dyakonov1971b,Hirsch1999,Sinova2015}, current-induced spin-orbit torques \cite{Manchon2008,Manchon2009,Manchon2015,Manchon2019}, magnetic skyrmions \cite{Bogdanov2001,Bogdanov2006,Muhlbauer2009,Yu2010}, and magnon thermal Hall effect \cite{Onose2010,Ideue2012,Hirschberger2015}. 

In magnetic materials, the elementary excitations are represented by spin-waves (SWs), which were first introduced by Bloch in 1929 with a spin wave theory in the Heisenberg model of ferromagnetism \cite{Bloch1930}. Then the spin wave theory was further developed to determine the ground state energy and excitation spectrum in ferro- and antiferromagnetism \cite{Holstein1940,Anderson1952,Kubo1952,Dyson1956,Oguchi1960}. From a quasiparticle point of view, SWs are collective quasiparticle excitations of the magnetic system, which can be understood as a coherent precession of localized magnetic moments \cite{Prabhakar2009}. Analogous to photons or phonons, quanta of SWs are referred to as magnons, which are bosonic quasiparticles. Essentially, the collective excitations of magnons can be ascribed to both the short-range exchange interaction (e.g. Heisenberg exchange interaction, Dzyaloshinskii–Moriya interaction) and the non-local exchange or long-range interaction (e.g. magnetic dipolar interaction). As a magnon is a `charge free' quasi-particle, it exhibits clear advantages over traditional electronic devices both low energy dissipation and long coherence length \cite{Cornelissen2015,Lebrun2018}, which renders magnons a promising alternative to electrons as information carriers \cite{Khitun2010,Chumak2014}. This gives rise to a new emerging research field, so-called magnonics (or magnon spintronics), which aims to deal with the excitation, propagation, control, and detection of magnons \cite{Kruglyak2006,Kruglyak2010,Serga2010,Lenk2011,Chumak2015,Barman2021}. Although magnonics is a young and developing research field, a flurry of research have unraveled various properties of magnons, such as generation \cite{Serga2004,Liu2007}, propagation \cite{Covington2002,Demidov2009}, reflection and refraction \cite{Stigloher2016,Zhuo2022}, interference \cite{Podbielski2006}, diffraction \cite{Birt2009}, and Doppler effect \cite{Demokritovtu2004,Vlaminck2008}. Hence, a mass of devices and concepts, including magnonic interferometer \cite{Hertel2004}, waveguides \cite{Sanchez2015}, multiplexors \cite{Vogt2014}, splitter \cite{Sadovnikov2015}, diodes \cite{Lan2015}, logic gates \cite{Jamali2013}, all-magnon logic circuits \cite{Khitun2010}, and neuromorphic computing \cite{Gartside2022,Chumak2022}, have been unearthed.
 
Over the past few decades, understanding and exploring the concept of topology in condensed matter physics is another hot topic. Following Ginzburg-Landau theory, phases of matter are described by microscopic order parameters, which characterize the internal structures of the physical system in terms of corresponding symmetries. And the abrupt changes of parameters are often associated with specific symmetry breaking during a phase transition \cite{Ginzburg1950}. But in some special cases such as the (fractional) quantum Hall states \cite{Thouless1982,Kalmeyer1987}, quantum spin liquids \cite{Wen1989,Wen2002}, topological insulators \cite{Qi2008}, and magnetic skyrmions \cite{Nagaosa2013}, order parameters can not be clearly characterized by symmetry breaking. A different classification paradigm so-called `topological order' as a quantum order was then introduced \cite{Thouless1982,Wen1985}, which defines a topological phase by a global topological index rather than by the local geometry \cite{Xiao2010,Ren2016}. Although the topological index in these systems is insensitive to smooth changes in a specific parameter space without any symmetry breaking, the topological order changes when the system passes through a quantum phase transition \cite{Kane2005,Hasan2010,Chiu2016}. 

More recently, the topological band theory has been established to discover and understand salient characteristics of topological states in a wide range of quantum  materials \cite{Bansil2016}, such as insulators, semimetals, superconductors, and superfluids \cite{Fu2007,Qi2011,Yan2012,Yan2017,Armitage2018}. One of the core principles in topological band theory is the connection between the topological invariants and the nontrivial topological phases, for instance, the Thouless–Kohmoto–Nightingale–den Nijs (TKNN) invariant or Chern number corresponds to the gapless boundary states in two dimensions electron system with time-reversal symmetry breaking \cite{Thouless1982,Kohmoto1985}. So far, studies of topological band theory have mostly concentrated on electron systems. In the last decade, there has been a growing interest in systems consisting of bosonic collective excitations, such as photons \cite{Lu2014,Khanikaev2017,Bansil2019}, phonons \cite{Süsstrunk2015,Zhang2018}, Cooper pairs \cite{Schnyder2015,Hasan2015,Sato2017}, excitons \cite{Wu2017,Kwan2021}, and magnons \cite{Murakami2017,McClarty2021,Bonbien2021,Li2021}. 

In the present review, we aim to provide an up-to-date survey on the topological aspects of magnonic systems from the point-of-view of theory and experiment, but do not wish to deliver an exhaustive overview of the vast field of magnon spintronics \cite{Barman2021}. There are several excellent reviews related to this topic \cite{McClarty2021,Bonbien2021,Li2021}. This review attempts to give a simple but detailed introduction to the current status of this topic, as we have tended to cite more recent literature in the following sections. However, it does not mean that this review could be considered an exhaustive review of this fast-evolving topic. It is organized as follows. In Section 2, we will first introduce some basic notions and necessary theoretical fundamentals on topological magnons, including the Berry phase, Chern number, Hall effect, and topological phase transition of magnons. In Section 3, topological phases of magnons are classified. We mainly focus on recent progress in magnon Chern insulators, high-order topological magnon insulators, $\mathbb{Z}_2$ topological magnon insulators, and topological magnon semimetals. In Section 4, candidate materials and artificial structures are summarized, which are the most successful platforms to observed topological magnons. In Section 5, we provide a short summary and outlook on this field. 

\section{Underpinnings of Topological Magnons}
In this section, we briefly outline the topological band theory in magnonic systems. We first review the linear spin-wave theory and then give the Berry phase and Berry curvature of magnons within the framework. Then, we introduce the topological invariant in a magnonic system---Chern number of magnons. Next, we introduce a semiclassical picture for understanding the dynamics of magnon wavepackets. We also discuss the Hall effect and the topological phase transition of magnons.
\subsection{Berry Phase and Chern Number of Magnons}
First, we introduce the linear spin-wave theory (LSWT) in magnetic systems with collinear magnetic moments (i.e., ferromagnets or collinear antiferromagnets) \cite{Colpa1978,Shindou2013}. Besides, we encourage the reader interested in deeper discussions of the LSWT for noncollinear magnetic systems, such as noncollinear antiferromagnets and skyrmions, to refer to these specialized articles \cite{Haraldsen2009,Toth2015,Mook2019}, as well as the nonlinear spin-wave theory for interacting magnonic systems with magnon-magnon interactions to refer to the articles \cite{McClarty2021,McClarty2019,Mook2021}. 

We consider the following generic spin Hamiltonian
\begin{equation}
		\mathcal{H} = \sum_{i,j}^L \sum_{m,n}^N \boldsymbol {S} _{i,m}^{\rm T} J_{ij}^{mn} \boldsymbol {S} _{j,n},
		\label{spinhamiltonian}
\end{equation} 
with spin operators $\boldsymbol {S}_{i,m}$ and $\boldsymbol {S} _{j,n}$, where indexes $i$ and $j$ run over the $L$ magnetic unit cells and $m$ and $n$ run over the $N$ magnetic sublattices in the magnetic unit cell. Magnetic interactions between two spin operators are comprised in $J_{ij}^{mn}$, including Heisenberg exchange interaction, Dzyaloshinskii-Moriya (DM) interaction, magnetostatic dipolar interaction, and so on. Then we express the spin operators in Eq. \eqref{spinhamiltonian} in terms of magnon creation operator $ \hat{b} ^{\dagger }_{i,m}  $ and annihilation operator $ \hat{b}_{i,m} $ by performing the Holstein-Primakoff transformations \cite{Holstein1940}
\begin{equation}
	\begin{gathered}
		S^{+}_{i,m}= \sqrt{2S- b^{\dagger}b_{i,m}}b_{i,m}, \\
        S^{+}_{i,m}= b^{\dagger }_{i,m}\sqrt{2S-b^{\dagger}b_{i,m}}, \\
        S^{z}_{i,m} = S-b^{\dagger }_{i,m}b_{i,m},
    \end{gathered}
	\label{HP}
\end{equation} 
where we introduce the magnon ladder operators $ S^{\pm}_{i,m}=S^{x}_{i,m} \pm iS^{y}_{i,m} $. In the low-temperature limit, the square roots can be expanded in powers of $1/\sqrt{S}$ when considering $2S \gg \langle n_{i,m} \rangle = \langle b^{\dagger}_{i,m}b_{i,m} \rangle$, because $n_{i,m}$ the number of thermally excited magnons, is small. After a Fourier transformation, we obtain the bilinear magnon Hamiltonian in the momentum space
\begin{equation}
		\mathcal{H}_{m}^{(2)} = \sum_{\boldsymbol k} \boldsymbol \Psi_{\boldsymbol k}^{\dagger} \mathcal{H}_{\boldsymbol k} \boldsymbol \Psi_{\boldsymbol k},
		\label{H2}
\end{equation} 
with the linear spin-wave matrix
\begin{equation}
		\mathcal{H}_{\boldsymbol k} = \begin{pmatrix}
			\mathbf {A}_{\boldsymbol k} & \mathbf {B}_{\boldsymbol k}\\
			\mathbf {B}_{-\boldsymbol k}^{\ast} & \mathbf {A}_{-\boldsymbol k}^{\ast}
		\end{pmatrix}
		\label{HSW}
\end{equation} 
and the vector boson operator
\begin{equation}
		\Psi_{\boldsymbol k}^{\dagger} = (b_{\boldsymbol k,1}^{\dagger},\cdots,b_{\boldsymbol k,N}^{\dagger},b_{-\boldsymbol k,1},\cdots,b_{-\boldsymbol k,N}).
		\label{basispsi}
\end{equation} 

\begin{figure}[t]
	\centering
	\includegraphics[width=0.46\textwidth]{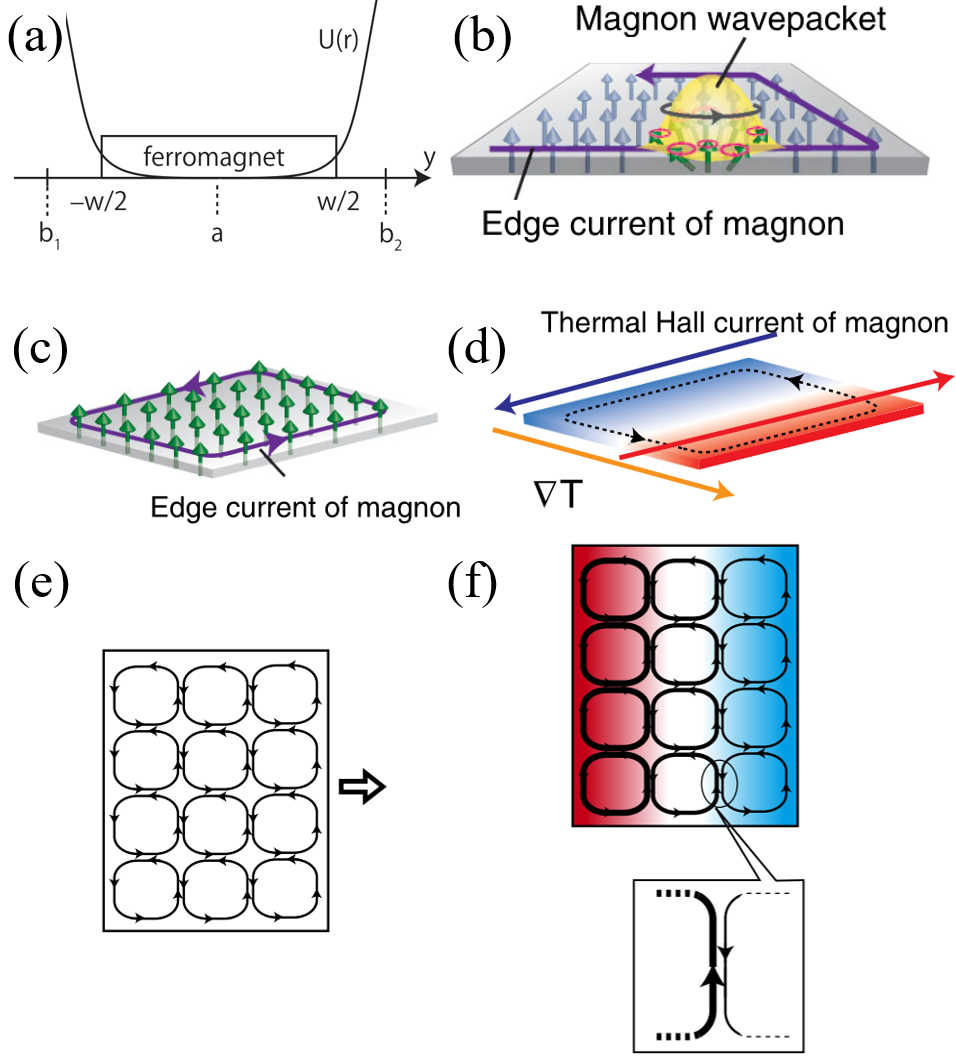}
	\caption{a) A confining potential in a ferromagnetic nanoribbon. Adapted with permission.\cite{Matsumoto2011b} Copyright 2011, American Physical Society. b) Self-rotation of a magnon wavepacket and a magnon edge current. c) Edge current of magnon in equilibrium. d) Under the temperature gradient, a finite thermal Hall current of magnon will appear. Adapted with permission.\cite{Matsumoto2011a} Copyright 2011, American Physical Society. e) In equilibrium, the magnon edge currents within the neighboring small regions cancel each other. f) Under the temperature gradient, a net transverse current appears as the magnon edge currents within the neighboring small regions do not cancel each other. Adapted with permission.\cite{Murakami2017} Copyright 2017, The Physical Society of Japan.}
	\label{fig:2-1}
\end{figure}
At first glance, the magnon Hamiltonian is similar to a Bogoliubov-de-Gennes Hamiltonian from superconductivity. The linear spin-wave matrix is diagonalized by performing a paraunitary Bogoliubov transformation
\begin{equation}
    \begin{aligned}
		E_{\boldsymbol k} & \equiv T_{\boldsymbol k}^{\dagger} \mathcal{H}_{\boldsymbol k} T_{\boldsymbol k} \\
        &= \mathrm{diag} (E_{\boldsymbol k,1},\cdots,E_{\boldsymbol k,N},E_{-\boldsymbol k,1},\cdots,E_{-\boldsymbol k,N}),
    \end{aligned}
	\label{eigenv}
\end{equation} 
and the magnon Hamiltonian reads
\begin{equation}
		\mathcal{H}_{m}^{(2)} = \sum_{\boldsymbol k} \sum_{n=1}^N E_{\boldsymbol k,n} \left ( b_{\boldsymbol k,n}^{\dagger}b_{\boldsymbol k,n}+\frac{1}{2} \right ).
		\label{H22}
\end{equation} 
where $E_{\boldsymbol k,n}$ is the $n$th band energy of magnon. In addition, the paraunitary Bogoliubov transformation must satisfy
\begin{equation}
		T_{\boldsymbol k}^{\dagger} G T_{\boldsymbol k} = G,
		\label{PBT}
\end{equation} 
where $G=\mathrm{diag} (1,\cdots,1,-1,\cdots,-1)$ with N values each of positive and negative one along the diagonal. Different from  the analogous unitary matrix to diagonalize a fermionic Hamiltonian, the paraunitary matrix $T_{\boldsymbol k}$ is not unitary in bosonic systems. Then the Berry curvature of magnons in the $n$th band is defined as \cite{Mook2019}
\begin{equation}\label{Berrycurvature}
	\Omega _{n \boldsymbol {k}}^{\mu\nu} = -2\sum_{m \ne n}^{2N} \mathrm{Im} \frac{( G T_{\boldsymbol k}^{\dagger} \partial _{\mu} \mathcal{H}_{\boldsymbol k} T_{\boldsymbol k} )_{nm} ( G T_{\boldsymbol k}^{\dagger} \partial _{\nu} \mathcal{H}_{\boldsymbol k} T_{\boldsymbol k} )_{nn}}{\left [ (G E_{\boldsymbol {k}})_{nm}-(G E_{\boldsymbol {k}})_{mm}  \right ]^{2}}.
\end{equation}
Finally, the Chern number for the nth magnonic bulk band is given by the integration of its Berry curvature over the Brillouin
zone (BZ)
\begin{equation}\label{Chernnumber}
	C_{n}=\frac{1}{2\pi }\int_{BZ} d\boldsymbol {k} \Omega _{n \boldsymbol {k}}.
\end{equation} 

\begin{figure}[t]
	\centering
	\includegraphics[width=0.42\textwidth]{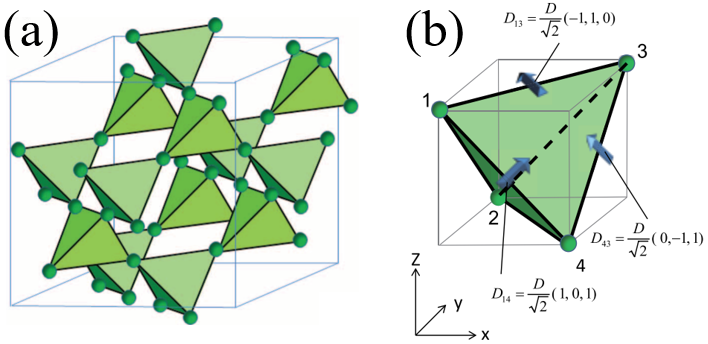}
	\caption{a) The crystal structure of Lu$_2$V$_2$O$_7$. The V$^{4+}$ sublattice is composed of corner-sharing tetrahedra. Adapted with permission.\cite{Onose2010} Copyright 2010, AAAS. b) The direction of the DM vector on each bond of the tetrahedron. Adapted with permission.\cite{Ideue2012} Copyright 2011, American Physical Society.}
	\label{fig:2-2}
\end{figure}
\subsection{Rotational Motion of Magnons Under Nonzero Berry Curvatures}\label{sec2.2}
Before discussing the Hall effect of magnons, let us first briefly describe the dynamics of a magnon wavepacket in a periodic magnonic system. When an external force is applied to an electron, it will undergo a transverse motion perpendicular to the external force, i.e., an intrinsic Hall effect such as the anomalous Hall effect \cite{Nagaosa2010} and spin Hall effect \cite{Sinova2015}. In semiclassical theory, when an electron wavepacket is localized both in real and momentum space, its dynamics can be described by the semiclassical equations of motion \cite{Xiao2010,Chang1996,Sundaram1999}. In analogy with this, Matsumoto and Murakami \cite{Matsumoto2011a,Matsumoto2011b} propose a similar framework for magnons, which gives the semiclassical equations of motion for the magnon wave packet as
\begin{gather}
	\dot{\boldsymbol {r}} = \frac{1}{\hbar}\frac{\partial E_{\boldsymbol {k},n}}{\partial \boldsymbol {k}} - \dot{\boldsymbol {k}} \times \Omega _{n \boldsymbol {k}} \label{semiclassicalequations1} \\ 
    \hbar \dot{\boldsymbol {k}} = -\nabla U(\boldsymbol {r}) \label{semiclassicalequations2}
\end{gather}  
where $U(\boldsymbol {r})$ is a slowly varying potential for the magnons in real space. Different from electrons, magnons are charge-neutral quasiparticles, thus they are immune to the Lorentz force from the external electric field. Analogous to the approach to describe the edge picture of the quantum Hall effect in electron systems with a confining potential \cite{Buttiker1988}, $U(\boldsymbol {r})$ can be regarded as a confining potential that changes from zero to infinity as the position $\boldsymbol {r}$ changes from inside the sample to the outside (see \Figure{fig:2-1}a). This confining potential $U(\boldsymbol {r})$ forbids the magnon wavepacket from running away from the sample, and its gradient $\nabla U(\boldsymbol {r})$ exerts a force on the wavepacket. From Eq. \eqref{semiclassicalequations1} we can see that there exists an edge magnon current in equilibrium due to the anomalous velocity term $\dot{\boldsymbol {k}} \times \Omega _{n \boldsymbol {k}}=-\nabla U(\boldsymbol {r})/\hbar \times \Omega _{n \boldsymbol {k}}$ induced by the large gradient of confining potential near the edge of the sample. Meanwhile, the magnon wavepacket also shows a self-rotation motion due to the Berry phase (see \Figure{fig:2-1}b). Hence, the angular momentum of the edge magnon current and that of the self-rotation motion give the total orbital angular momentum of magnons, which has been demonstrated with linear response theory. 
\begin{figure*}[t]
	\centering
	\includegraphics[width=0.9\textwidth]{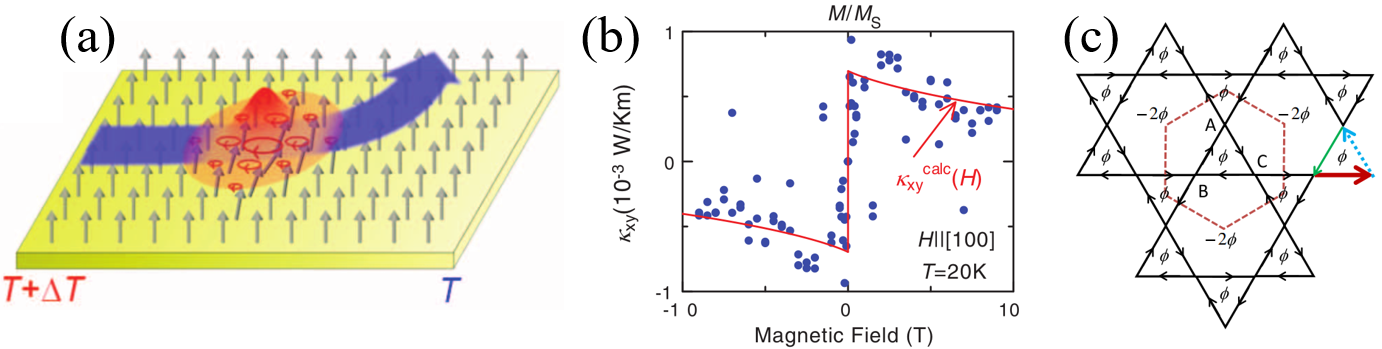}
	\caption{a) Thermal Hall effect of magnons. A magnon wavepacket moving from the hot to the cold side experiences a transverse motion. b) Thermal Hall conductivity as a function of the external magnetic field. The red solid line indicates the theoretical results. Adapted with permission. \cite{Onose2010} Copyright 2010, AAAS. c) Fictitious magnetic flux due to the DM interaction in the kagome lattice (the [111] plane of the pyrochlore lattice). Adapted with permission. \cite{Katsura2010} Copyright 2011, American Physical Society.}
	\label{fig:2-3}
\end{figure*}

In addition, a nonzero Berry curvature is also necessary for the anomalous velocity to generate an edge magnon current and the self-rotation motion of the magnon wavepacket. Hence, similar to the spin Hall effect in electronic systems, it requires some kind of “spin–orbit interaction” to ensure the Berry curvature of magnons is nonzero. To the best of the authors’ knowledge, all of the established materials or models that exhibit a nonzero Berry curvature of magnons mostly rely on the following three interactions: 

1) The antisymmetric exchange interaction---DM interaction. It is natural because the DM interaction itself originates from the spin-orbit interaction in first-order perturbation theory, when the inversion symmetry of the system is broken \cite{Dzyaloshinsky1958,Moriya1960}. For example, the nonzero Berry curvature of magnons in the ferromagnetic Mott-insulator Lu$_2$V$_2$O$_7$ \cite{Onose2010,Ideue2012}, whose spin-$1/2$ V$^{4+}$ ions are composed of corner-sharing tetrahedra forming a pyrochlore structure (see \Figure{fig:2-2}a). There is a nonzero DM interaction with DM vectors perpendicular to the vanadium bond and parallel to the face of the surrounding cube (see \Figure{fig:2-2}b), because the midpoint between any two apices of a tetrahedron is not a center of inversion symmetry in the pyrochlore structure. Since the total DM vector of the six bonds sharing the same site is zero, the ground state is a collinear ferromagnet below Curie temperature ($T_C\approx 70 \mathrm{K}$). 

2) The magnetic dipolar interaction for the magnetostatic SWs with a sufficiently long wavelength ($\thicksim \mu \mathrm{m}$) in a two-dimensional magnetic thin film \cite{Matsumoto2011a,Matsumoto2011b,Matsumoto2014} or in some artificial magnonic crystals \cite{Shindou2013b,Shindou2013c,Shindou2014}. In this case, the propagation of magnons or SWs is dominated by the long-range magnetic dipolar interaction rather than the short-range exchange interactions. The magnetic dipolar interaction depends both on the orientations and the relative positions of the magnetic moments, hence it plays the role of the spin–orbit interaction and brings about a nonzero Berry curvature. Moreover, the formalism of the magnon Berry curvature due to the magnetic dipolar interaction is quite different from that due to the DM interaction, because magnons are sensitive to the sample shape in the former case. 

3) The magnon-magnon interaction. Beyond the linear spin wave theory in terms of quadratic order boson operators, magnon-magnon interactions become important at higher temperatures, which can also be treated in non-linear spin wave theory (a perturbation theory) by analyzing higher-than-bilinear contributions (such as the cubic and quartic terms) during the Holstein–Primakoff transformation. A few existing studies show that magnon-magnon interactions renormalize the magnon energy bands as an origin of nontrivial magnon topology and cause detrimental lifetime broadening effects \cite{Chernyshev2009,Chernyshev2015,Chernyshev2015b,Chernyshev2016,McClarty2018,Rau2019,Mook2022}. Since the interaction-induced self-energy is non-Hermitian and the magnon band gaps or crossings occur at finite energy, non-Hermitian magnon topology could be expected with topologically protected exceptional points \cite{McClarty2021,McClarty2019,Mook2021}. 

In addition, the detailed summary of other origination for a nonzero Berry curvature is not given here, such as weak ferromagnetism with a nonzero scalar spin chirality or an external magnetic field \cite{Hirschberger2015b,Schutte2014,Owerre2017b,Cookmeyer2018}.

Under a temperature gradient, the transverse magnon current appears. This phenomenon can be understood in the following semiclassical picture \cite{Murakami2017}: A magnet can be divided into a lot of small regions, and meanwhile, there will exist edge currents along the edges of each individual small region due to the confining potential at the edge (see \Figure{fig:2-1}e). Since the size of each region is very small, the difference between neighboring regions should be negligible. Thus the internal edge currents in each small region cancel each other, leaving behind the magnon current along the edge of the magnet (see \Figure{fig:2-1}c). When the temperature gradient is present, the magnon edge currents in each region are different and do not cancel between neighboring regions (see \Figure{fig:2-1}f), then a net transverse magnon current is generated (i.e. the thermal Hall effect of magnons, see \Figure{fig:2-1}d).

\subsection{Hall Effect of Magnons}
\subsubsection{Thermal Hall Effect of Magnons}
From \eqref{semiclassicalequations1}, the semiclassical equation of motion for magnons, magnon wavepackets will have a transverse velocity (i.e. so-called anomalous velocity) perpendicular to the external force ($\hbar \dot{\boldsymbol {k}}$). That is the Hall effect of magnons. In the case of the Hall effect of photons, a spatial gradient of the refractive index often plays the role of an external force to supply the anomalous velocity \cite{Onoda2004,Bliokh2015}. For magnons, a temperature gradient could serve as an external force, which induces the thermal Hall effect of magnons also known as magnon thermal Hall effect (see \Figure{fig:2-3}a). When a finite thermal Hall current is driven by a longitudinal temperature gradient in a two-dimensional magnet, the thermal Hall conductivity is given by
\begin{equation}\label{THC}
	\kappa _{xy} = -\frac{k_{B}^{2}T}{ \hbar}\sum_{n,\boldsymbol {k}} c_{2}\left [ \rho^{B} \left ( E _{\boldsymbol {k},n} \right )   \right ]  \Omega _{n \boldsymbol {k}}^{z},  	
\end{equation}
where $ \rho^{B} \left ( E _{\boldsymbol {k},n} \right )=\left(  e^{E _{\boldsymbol {k},n}/k_{B}T }-1  \right) ^{-1} $ is the Bose-Einstein distribution. The weighting function is given by $ c_{2}(\rho^{B})=\left ( 1+\rho^{B} \right ) \ln^{2} {\frac{1+\rho^{B}}{\rho^{B}}}-\ln^{2} {\rho^{B}} -2\mathrm{Li}_{2}\left ( -\rho^{B} \right ) $, with $ \mathrm{Li}_{2}\left( \rho^{B} \right) $ being the polylogarithm function. It shows that thermal Hall conductivity is completely determined by the Berry curvature of magnons with a temperature distribution. 
\begin{figure}[t]
	\centering
	\includegraphics[width=0.42\textwidth]{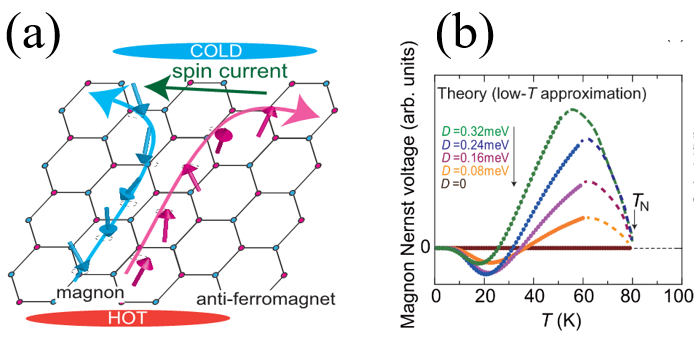}
	\caption{a) The spin Nernst effect of magnons in honeycomb antiferromagnets. b) Temperature dependence of the spin Nernst signal  for selected DM interaction values in  monolayer MnPS$_3$. Adapted with permission. \cite{Shiomi2017} Copyright 2017, American Physical Society.}
	\label{fig:2-4}
\end{figure}

\begin{figure}[t]
	\centering
	\includegraphics[width=0.44\textwidth]{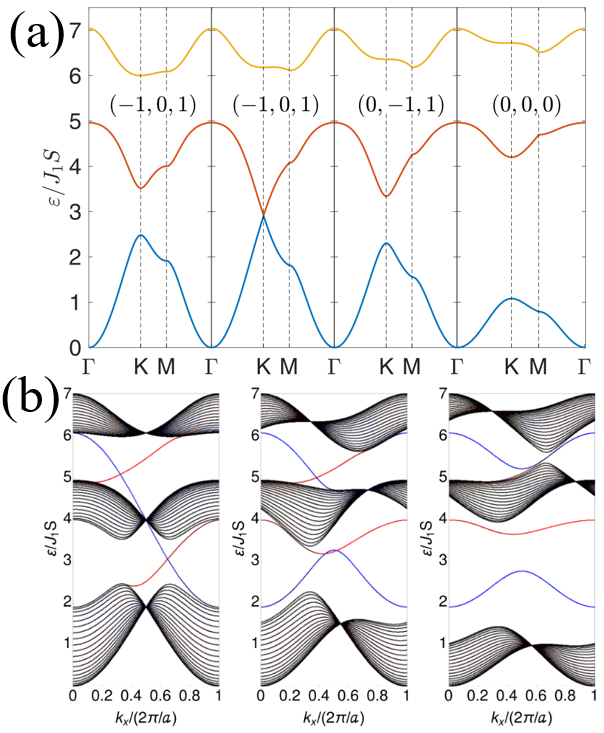}
	\caption{a) Topological phase transitions of magnon band structures in kagome ferromagnets with selected lattice deformation parameter, from left to right panel $\delta=$0, 0.05, 0.1, 0.2. b) Chiral magnonic edge states in a kagome lattice, from left to right panel $\delta=$0, 0.1, 0.18. Adapted with permission. The black lines are the bulk bands and the blue/red lines are the edge states. Adapted with permission. \cite{Zhuo2021} Copyright 2021, American Physical Society.}
	\label{fig:2-5}
\end{figure}
The theory of the thermal Hall effect of magnons was first established by Katsura et al. \cite{Katsura2010}, who computed the thermal Hall conductivity using the Kubo formula in a kagome lattice ferromagnet with the DM interaction. In their theory, the DM interaction imprints a lattice geometrical phase for magnons, so-called fictitious magnetic flux (see \Figure{fig:2-3}c), which is important to avoid cancellation of the effect of phase factor in the unit cell. Subsequently, Onose et al. \cite{Onose2010} observed the thermal Hall effect in the ferromagnetic Mott-insulator Lu$_2$V$_2$O$_7$ as mentioned above. In the experiment, they measured the thermal Hall conductivity and compared their data with theoretical results (see \Figure{fig:2-3}b). When switching the magnetic field, the thermal conductivity shows a sign reversal. This confirms that the thermal Hall effect originates from the magnons rather than phonons. Up to now a growing number of studies have been investigating the DM interaction driven thermal Hall effect of magnons on specific lattice geometries, including honeycomb \cite{Owerre2016,Li2021b,Zhang2021,Neumann2022}, triangular \cite{Chen2022}, kagome \cite{Hirschberger2015,Chisnell2015,Mook2014,Laurell2018,Mook2019,Zhuo2021,Zhuo2022b}, and Lieb lattice \cite{Cao2015,Pires2022}. Significantly, the thermal Hall effect of magnons could be absent in some magnetic systems despite the presence of the DM interaction \cite{Ideue2012}. Mook et al. \cite{Mook2019} proposed that a broken effective time-reversal symmetry and a magnetic point group compatible with ferromagnetism are two necessary requirements for the DM interaction driven thermal Hall effect of magnons. Furthermore, a few studies also show the thermal Hall effect of magnons in some specific spin configurations without the DM interaction \cite{Owerre2017b,Kim2019,Fujiwara2022,Kawano2019,Kawano2019,Rosales2019,Albarracin2021}, a two-dimensional magnetic thin film due to the magnetic dipolar interaction \cite{Matsumoto2011a,Matsumoto2011b,Matsumoto2014}, and a Skyrmion lattice \cite{Hoogdalem2013,Mook2017,Kim2019b,Carnahan2021,Akazawa2022}. In the last case, the fictitious magnetic fields due to the equilibrium magnetic texture lead to the thermal Hall effect of magnons.  
\begin{figure}[t]
	\centering
	\includegraphics[width=0.44\textwidth]{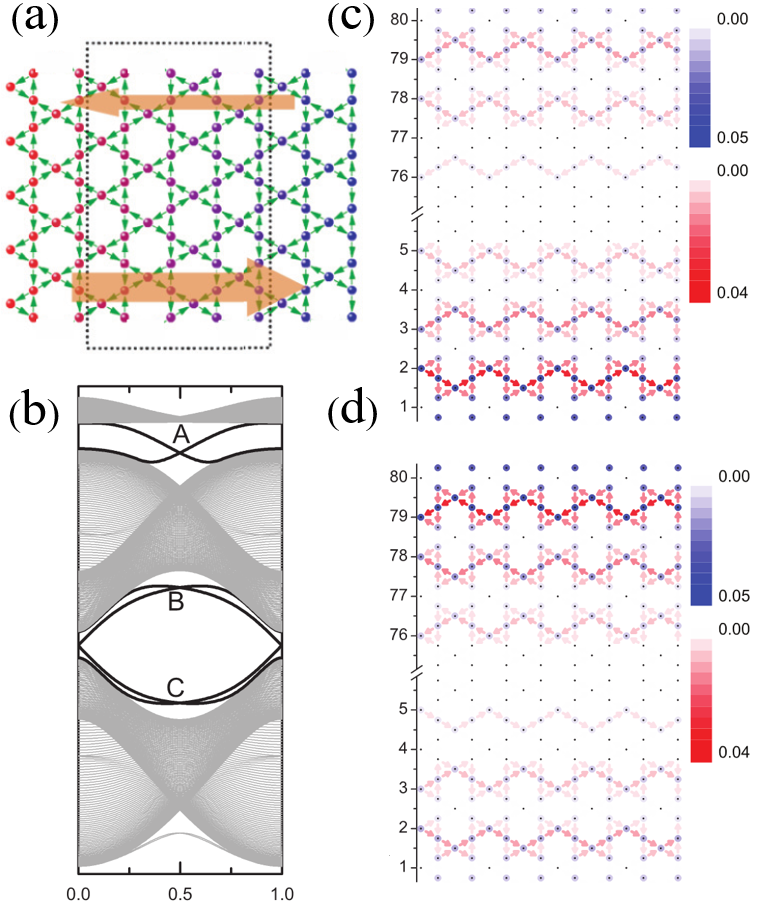}
	\caption{a) The two-dimensional kagome-lattice nanoribbon. The temperature on the left side ($T_L$) is higher than that on the right side ($T_R$). The two big arrows show the magnitudes and directions of the magnon edge currents. b) The magnon band structure of the kagome-lattice nanoribbon. The local energy current and density of state for edge magnon transport under a temperature gradient in c) $T_L>T_R$ and d) $T_L<T_R$. The color of the arrows and dots indicates the magnitude of the local current and density of states, respectively. Adapted with permission. \cite{Zhang2013} Copyright 2013, American Physical Society.}
	\label{fig:3-1}
\end{figure}

\subsubsection{Spin Nernst Effects of Magnons}
The spin Nernst effect describes a transverse pure spin current as a response to a temperature gradient, which has been observed in both electronic \cite{Sheng2017} and magnonic \cite{Shiomi2017} systems. The spin Nernst effect of magnons (i.e. the magnon spin Nernst effect) can be equivalent to an antiferromagnetic analog of the thermal Hall effect of magnons in ferromagnetic insulators, that two magnon currents with opposite spins flow in opposite transverse directions under a longitudinal temperature gradient (see \Figure{fig:2-4}a). This effect could also be viewed as the magnonic version of the spin Hall effect driven by the spin Berry curvature of magnons. Cheng et al. \cite{ChengR2016} and Zyuzin et al. \cite{Zyuzin2016} theoretically demonstrated the spin Nernst effect of magnons in a collinear honeycomb antiferromagnet independently. Significantly, the spin Nernst effect coefficient shows a sign change due to the sign flip of the spin Berry curvature across the von Hove singularities as seen in \Figure{fig:2-4}b. Different from the thermal Hall effect of magnons requiring certain symmetries breaking, the spin Nernst effect of magnons is much more robust in collinear antiferromagnets, which can be driven by the DM interaction and even exists in systems with both time-reversal symmetry and inversion symmetry as long as a nonzero spin Berry curvature is present. In both cases the thermal Hall effect of magnons is absent. Especially in the former case, the spin Nernst effect coefficient changes sign with the reversal of the Ne\'el vector. In addition, the spin Nernst effect of magnons is also widely predicted in kagome ferromagnets \cite{Kovalev2016}, collinear honeycomb ferrimagnets \cite{Park2020}, and noncollinear kagome antiferromagnets \cite{Mook2019b,LiB2020} with the Rashba-like (in-plane) DM interaction, bilayer two-dimensional van der Waals magnets \cite{Go2022}, antiferromagnetic skyrmion crystals \cite{Diaz2019}, and even in paramagnets \cite{ZhangY2018}. Moreover, Kondo and Akagi \cite{Kondo2022} derived the formula for the spin Nernst effect of magnons in the nonlinear response regime, the so-called nonlinear magnon spin Nernst effect. This effect originates from a dipole moment of the Berry curvature (i.e. the Berry curvature dipole) of magnons in the crystal momentum space when the inversion and rotational symmetries in a system are broken even without the DM interaction. They confirmed that the nonlinear magnon spin Nernst effect could exist in the square lattice antiferromagnets with bond dependences of the nearest-neighbor exchange interaction, and collinear antiferromagnets in the honeycomb or diamond lattice under pressure. 

To date, there is only one experimental observation of the spin Nernst effect of magnons in a thin-film MnPS$_3$ \cite{Shiomi2017}. In their experiment, a non-monotonic temperature dependence of the spin Nernst effect signal detected by voltages through the inverse spin Hall effect is indeed observed. But the sign reversal of the spin Nernst effect coefficient has not been reported, because the thermoelectric voltage could not be unambiguously separated from the inverse spin Hall voltage. The dependence of the spin Nernst effect signal on a perpendicular magnetic field was also not measured in the experiment. More carefully designed measurements, such as using optical detection instead of electronic detection, are needed to identify the theoretical predictions.  

\begin{figure}[t]
	\centering
	\includegraphics[width=0.44\textwidth]{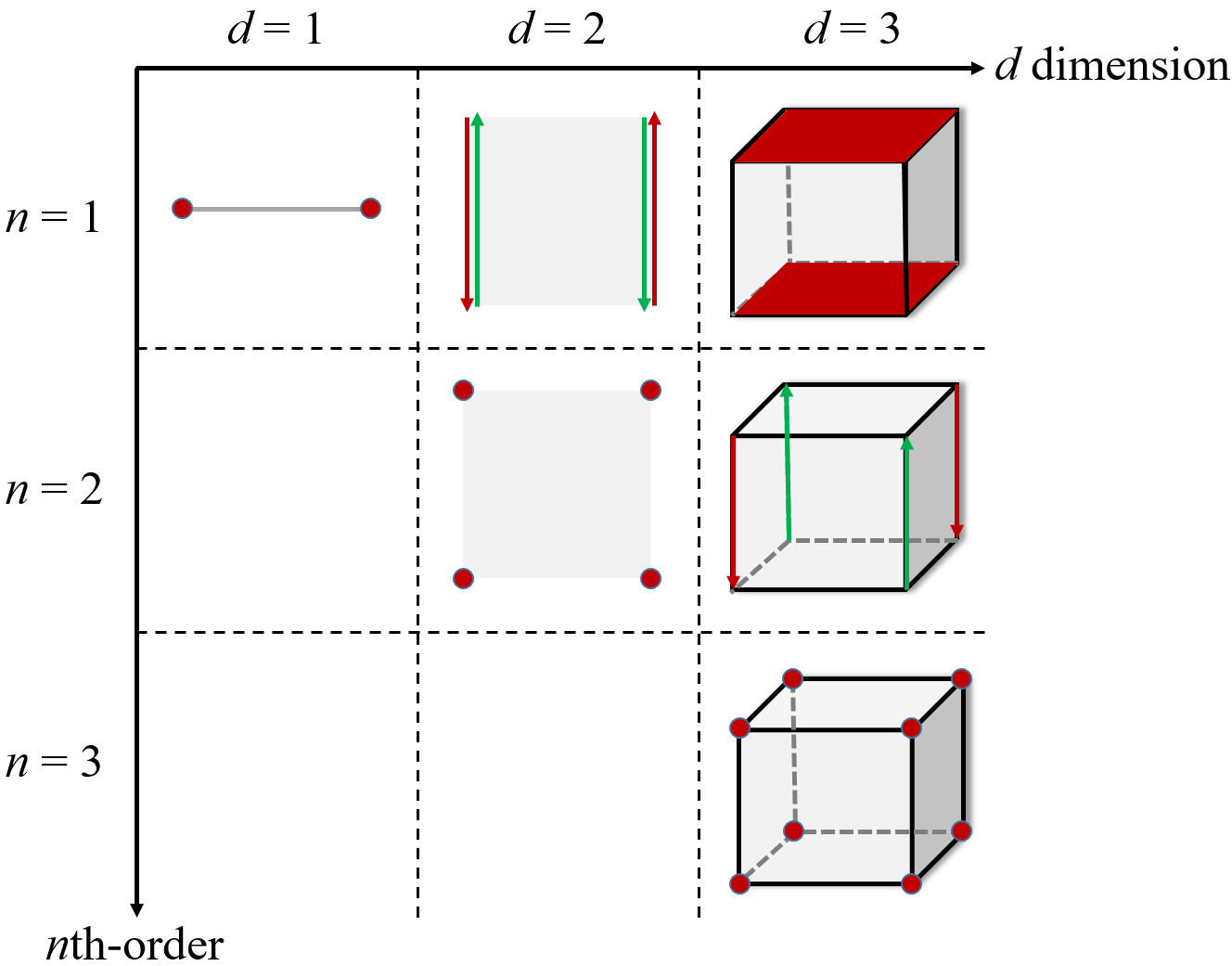}
	\caption{A schematic diagram of nth-order topological phases in d dimensions. The first line with $n=1$ corresponds to conventional or first-order topological insulators with gapless states, including corner states ($d=1$), edge states ($d=2$), and surface states ($d=3$). The lines with $n \geq 2$ correspond to higher-order topological insulators with gapless states, including corner states ($n=d=2$ or 3) and hinge states ($n=2$, $d=3$).}
	\label{fig:3-2}
\end{figure}

\begin{figure*}[t]
	\centering
	\includegraphics[width=0.9\textwidth]{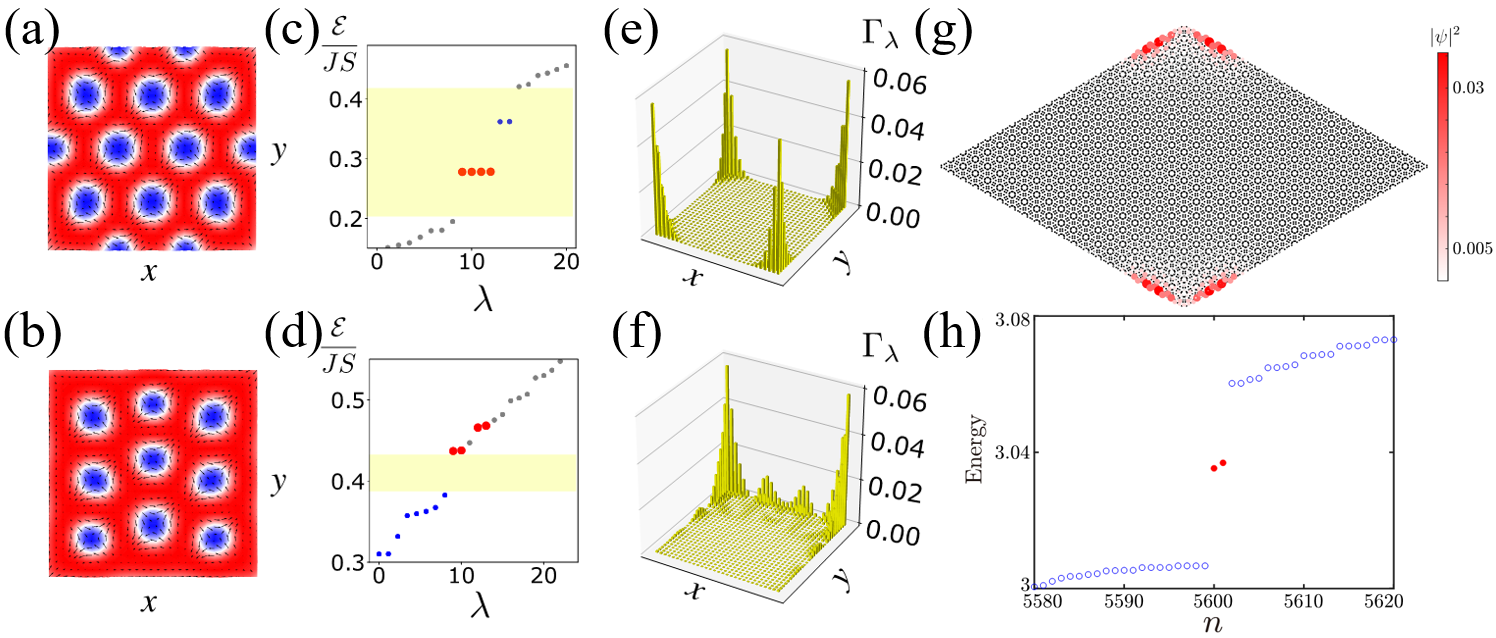}
	\caption{a) and b) Confined antiskyrmion crystals at different external magnetic fields. c) and d) Magnon spectrum showing corner states (red dots) and trivial bound states (blue dots) localized at the fractional antiskyrmions. e) and f) Probability density of the corner states. Adapted with permission. \cite{Hirosawa2020} Copyright 2020, American Physical Society. g) The spatial distribution of the probability density. h) Magnon energy spectrum. Red dots mark the in-gap magnonic corner states. Adapted with permission. \cite{HuaC2023} Copyright 2023, American Physical Society.}
	\label{fig:3-3}
\end{figure*}
\subsection{Topological Phase Transition of Magnons}
Different from the familiar phase transitions described by the Ginzburg-Landau theory such as the liquefaction of a solid, topological phase transition manifests itself from the changes in the topology of the bulk band structure. Topological phase transitions have been observed experimentally in topological insulators, which shows the evolution of the band structure with a band gap closing and reopening \cite{Bernevig2006,XuS2011,Sato2011}. In magnonic systems, topological phase transitions of magnons can be induced by tuning the magnon band structure using the magnetic field, temperature, magnon-phonon coupling, strain, or pressure \cite{Zhuo2021,Kim2019,Owerre2017,LuY2021,ZhangS2020,Go2019,Mook2014b,Moulsdale2019,Li2021b}. An example is given in \Figure{fig:2-5}a, that the nontrivial band gap between the two acoustic magnon branches closes around $\delta=0.05$, then reopens and becomes trivial as $\delta$ increasing under the strain \cite{Zhuo2021}. In this process, two chiral edge states in the gap go from nontrivial to trivial (see \Figure{fig:2-5}b). Similar results are found in honeycomb ferromagnets \cite{Li2021b}, where topological phase transitions accompanied by the sign reversal of the thermal Hall conductivity can be induced via tuning temperature or Zeeman field.  

\section{Classification of Topological Magnon Phases}
Before discussing topological magnons, let us first describe the differences between topological magnon insulators and topological (electron) insulators. In electronic systems, electrons can not flow on the surface or inside a conventional insulator due to the large band gap between the valence band and the conduction band, which mostly forbids electrons in the valence band jumping to the conduction band. Although topological insulators still have band gaps, there are edge states in the gaps supplying some unimpeded channels on the surfaces or boundaries for electrons flowing leading to the currents on the surface or boundary of the sample. But the interior of a topological insulator remains insulating. Since magnons are bosonic quasiparticles, all the magnon bands including the bulk bands and topological edge states contribute to the transport properties. Thus, a (topological) magnon insulator is never a ”true” insulator. As a matter of fact, the strict definition of the topological magnon insulator should be a magnon insulator has both bulk magnon bands and topological edge states, but the contributions to the transport properties from the edge states should absolutely dominate them from the bulk bands even being ignored. Then, recent extensive efforts have been paid to find ways to realize topological magnons. To date, a magnonic version of Chern insulators, high-order topological insulators, $\mathbb{Z}_2$ topological insulators, and topological semimetals has been proposed. In this section, we will systematically review these types of topological magnons. 

\begin{figure*}[t]
	\centering
	\includegraphics[width=0.9\textwidth]{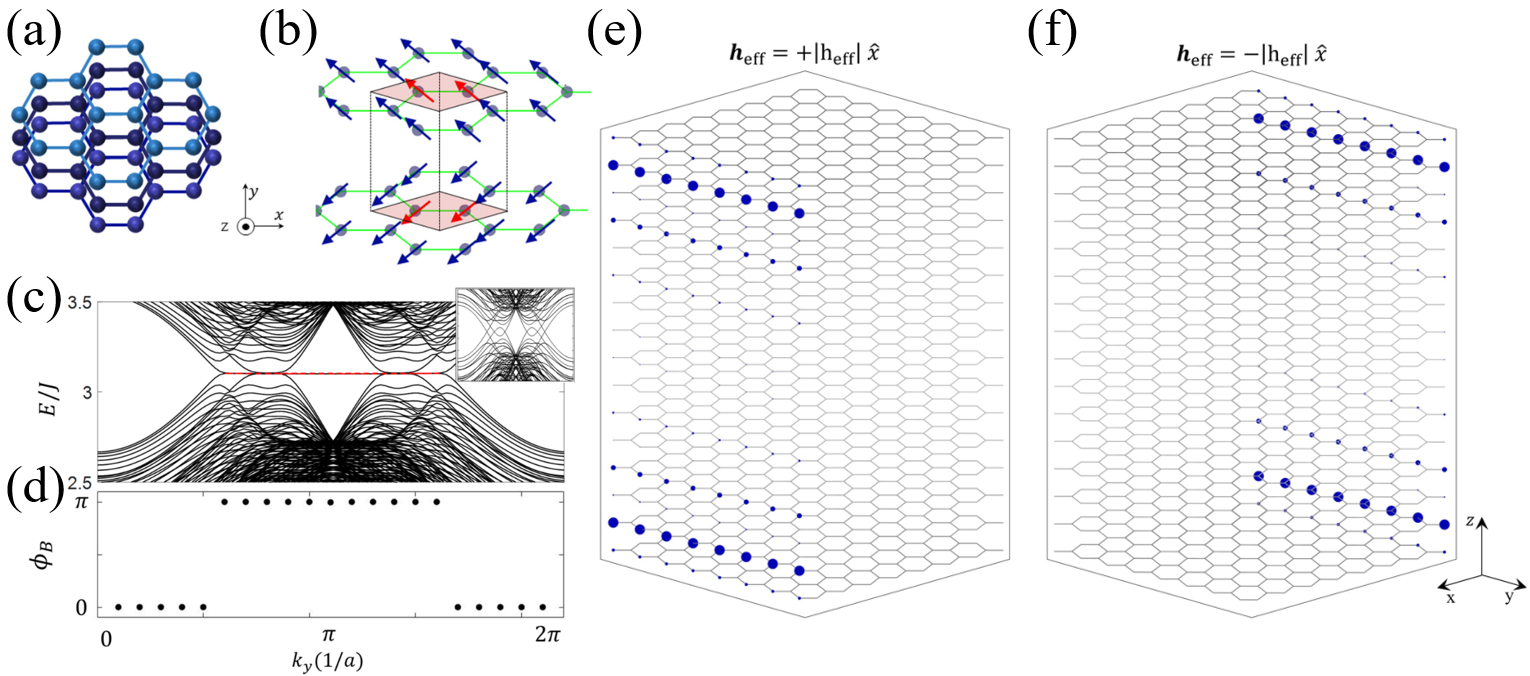}
	\caption{a) Top view of the atomic configurations in the vertically stacked monolayer honeycomb lattices. b) Noncollinear magnetic order under an external magnetic field. c) The magnon band structure as a function of $k_y$ with the open boundary condition along the $x-z$ direction. The red line in the middle of the bands marks the magnonic hinge modes. Inset: the magnon band structure with the periodic boundary condition along the $z$ direction. d) The non-Abelian Berry phase as a function of $k_y$. e) and f) The wave function of the magnonic hinge modes under opposite direction external magnetic fields. The periodic boundary condition is imposed along the $y$ direction. The size of the blue dot indicates the amplitude of the wave functions. Adapted with permission. \cite{Park2021} Copyright 2021, American Physical Society.}
	\label{fig:3-4}
\end{figure*}
\subsection{Magnon Chern Insulators}
As explained in Section 2.2, there exist magnon edge currents along the boundary of magnets due to both confining potentials and nonzero Berry curvatures in equilibrium. Under a temperature gradient, a net transverse current is generated by the temperature difference between the neighboring small regions when nonzero Berry curvatures are present. In this case, the magnon edge currents essentially originate from the topologically protected edge states in the magnon band structure, i.e. the topology of the topological magnon insulator. This is the magnon Chern insulator and we shall refer to it simply as the "topological magnon insulator". Zhang et al. \cite{Zhang2013} proposed the first topological magnon insulator in a ferromagnetic insulator with the DM interaction as shown in \Figure{fig:3-1}a. From the magnon band structure in \Figure{fig:3-1}b, we can clearly see the magnon edge states in the gaps. As shown in \Figure{fig:3-1}c and \Figure{fig:3-1}d, we can find that the magnon currents prefer to flow along one edge changing with the direction of the temperature gradient, which reflects the chirality of the magnon edge states. That's because the propagation directions (the red arrows) are determined by the directions of DM vectors (they give the signs of the Berry curvatures), and at the same time the magnons need to carry energy from the hot side to the cold one following the second law of thermodynamics. It is worth noticing that there are small bulk magnon currents inside the nanoribbon, although the currents mainly localize around two edges. It reflects that a topological magnon insulator is not a perfect topological insulator, where magnons in bulk bands can still transmit. In the meantime, Shindou et al. \cite{Shindou2013} proposed a magnonic topological insulator in a magnonic crystal, which provides topologically protected chiral edge states for magnetostatic spin waves due to the dipolar interaction. In these edge states, the SWs propagate in a unidirectional way without backward scatterings. So far, the topological magnon insulators have been theoretically investigated in kagome \cite{Zhuo2021,Zhuo2022b,Mook2014,Mook2014b,Chernyshev2016} and honeycomb \cite{Owerre2016,Li2021b,Owerre2016b,Owerre2016c,Owerre2016d} lattice in ferromagnetic systems with the DM interaction. Mook et al. \cite{Mook2014b} present the bulk-boundary correspondence using a Green function renormalization technique in the topological magnon insulator. Their results explain the sign of the transverse thermal Hall conductivity regarding topological edge states and their propagation direction. 

Besides, this topological magnon phase has been extended to various antiferromagnetic systems like canted collinear honeycomb lattice antiferromagnets \cite{Owerre2017c,Neumann2022}, canted noncollinear triangular lattice antiferromagnets \cite{Kim2019}, and noncollinear kagome antiferromagnets \cite{Owerre2017,Mook2019,Owerre2017d,Laurell2018}. In these systems, the canting of spin configurations due to an external magnetic field or a weak in-plane DM interaction gives rise to weak ferromagnetism, where a finite scalar spin chirality $\boldsymbol {S}_{i} \cdot \left ( \boldsymbol {S}_{j} \times \boldsymbol {S}_{k} \right ) $ can also produce the nontrivial topological magnon edge states even in the absence of DM interaction \cite{Katsura2010}. Corresponding to the thermal Hall effect of magnons in antiferromagnets, the two necessary demands for a topological magnon insulator are broken effective time-reversal symmetry and a magnetic point group compatible with ferromagnetism \cite{Mook2019}. 

\begin{figure*}[t]
	\centering
	\includegraphics[width=0.9\textwidth]{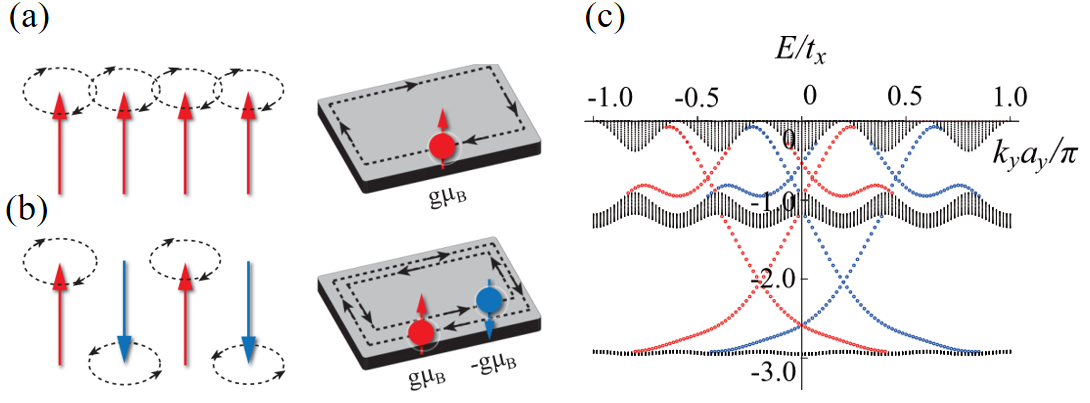}
	\caption{a) Schematic representation of spin excitations (Left) and edge magnon states (Right) in a ferromagnet. b) Schematic representation of spin excitations (Left) and edge magnon states (Right) in an antiferromagnet. c) Plots of the magnonic band structure. The up magnon edge states ($\sigma=1$) are in red while down magnon edge states ($\sigma=-1$) are in blue. Adapted with permission. \cite{Nakata2017} Copyright 2017, American Physical Society.}
	\label{fig:3-5}
\end{figure*}
\subsection{High-order Topological Magnon Insulators}
According to the topological connection between the bulk and boundary or namely the bulk-boundary correspondence \cite{Hasan2010,Qi2011,Bansil2016}, a $d$-dimensional $n$th order topological insulator has $\left(d-n \right)$ dimensional topologically protected gapless states $\left(d \leq n \right)$ as shown in \Figure{fig:3-2}. For example, a conventional or first-order $\left(n=1 \right)$ topological insulator in two dimensions $\left(d=2 \right)$ shows one-dimensional topological edge states in the band gaps (see Section 3.1). Recently, another extension of the topological insulator family, the so-called higher-order $\left(n \geq 2 \right)$ topological insulators, has become one of the cutting-edge research areas in condensed matter physics \cite{ZhangF2013,Benalcazar2014,Benalcazar2017,Benalcazar2017b,Langbehn2017,Song2017,Schindler2018}. Different from first-order topological insulators, higher-order topological insulators support lower-dimensional boundary signatures, which host zero-dimensional corner states $\left(n = d \geq 2 \right)$ and/or one-dimensional hinge modes $\left(n = d-1=2 \right)$ as shown in \Figure{fig:3-2}. In spite of a few experimental observations of the higher-order topological insulator in electronic materials  \cite{Schindler2018b,Choi2020,Aggarwal2021}, it has been extensively realized in various artificial materials or systems, such as electric circuits \cite{Imhof2018,LiuS2019,ShangC2022}, photonic \cite{Noh2018,Peterson2018,Mittal2019,Hassan2019}, acoustic \cite{XueH2019,NiX2019,ZhangZ2019,QiuH2021,DuJ2022,QiY2020}, and mechanical \cite{Garcia2018,FanH2019,WuX2021,Wakao2020,WuQ2020} metamaterials. 

In Section 3.1, we have discussed magnon Chern insulators in two-dimensional magnets, which can be viewed as first-order topological magnon insulators hosting one-dimensional topological edge states. In recent years, the concept of higher-order topological insulators has been similarly introduced into magnonic systems. A first example is given by Sil and Ghosh \cite{Sil2020}, who propose a second-order topological magnon insulator with localized magnonic corner states in two-dimensional breathing kagome ferromagnets. Then, Hirosawa et al. \cite{Hirosawa2020} uncovered that two-dimensional antiskyrmion crystals (see \Figure{fig:3-3}a and b) can also be used to realize a second-order topological magnon insulator, whose hallmark signatures are robust magnonic corner states. Tuning an external magnetic field can induce the self-assembly of fractional antiskyrmions along the edges of the sample (see \Figure{fig:3-3}c), which carry fractional topological charges allowing the emergence of corner localized magnonic edge states (red dots in \Figure{fig:3-3}e). Despite being topologically trivial bound states in the gap (blue dots in \Figure{fig:3-3}e), they locate inside the fractional antiskyrmions far away from the corners. In the case of the absence of fractional antiskyrmions as shown in  \Figure{fig:3-3}b, there are four significant edge modes near the corners (see \Figure{fig:3-3}f), which spread over the boundaries and then flow into the bulk of the sample due to mixing with bulk modes (see \Figure{fig:3-3}d). Another example is from Hua et al. \cite{HuaC2023}, where they show that twisted bilayer honeycomb ferromagnets can be used to realize second-order topological magnon insulators with magnonic corner states as shown in \Figure{fig:3-3}g. In the magnon energy spectrum plotted in \Figure{fig:3-3}h, it is found that two in-gap states (red dots) reside in the energy gap. These higher-order topological edge states strongly depend on the interlayer ferromagnetic exchange coupling. Their first-principles calculations show that a $\theta=21.78^\circ$ twisted bilayer van der Waals magnet, such as Chromium triiodide (CrI$_3$), could be the candidate material as experimental realizations of their theoretical model. 

However, since both the magnon Chern insulator in Section 3.1 and the second-order topological magnon insulator with magnonic corner states are realized in two-dimensional magnets, they are not suitable to be applied to current information technology tending to be three-dimensional integration. Recently, a three-dimensional second-order topological magnon insulator with magnonic hinge modes has been proposed by Park et al. \cite{Park2021} in vertically stacked honeycomb magnets with a noncollinear magnetic order due to the $x$-directional external magnetic field as shown in \Figure{fig:3-4}a and b. \Figure{fig:3-4}c shows the magnon band structure with an open boundary condition along the $z$ direction. A pair of in-gap states (red dashed line) emerges between $K$ and $K'$ points, which localize at the corner of the $x$-$z$ plane and show the nature of the magnonic hinge modes (see \Figure{fig:3-4}e and f). \Figure{fig:3-4}d shows the non-Abelian Berry phase. It is quantized and equal to $\pi$ where the hinge modes reside. This quantized Berry phase  gives rise to $\mathbb{Z}_2$-topological protected hinge mode. Most interestingly, different from the conventional electronic hinge modes, the magnonic hinge modes here localize only at the two corners of one side surface. And the localization of the hinge modes switches as the direction of the magnetic field is reversed. Alternatively, Mook et al. \cite{Mook2021b} proposed a second-order topological magnon insulator with magnonic hinge modes in vertically stacked honeycomb ferromagnets. Since ferromagnetism naturally breaks the time-reversal symmetry, the magnonic hinge modes are chiral without backscattering. 

\begin{figure*}[t]
	\centering
	\includegraphics[width=0.9\textwidth]{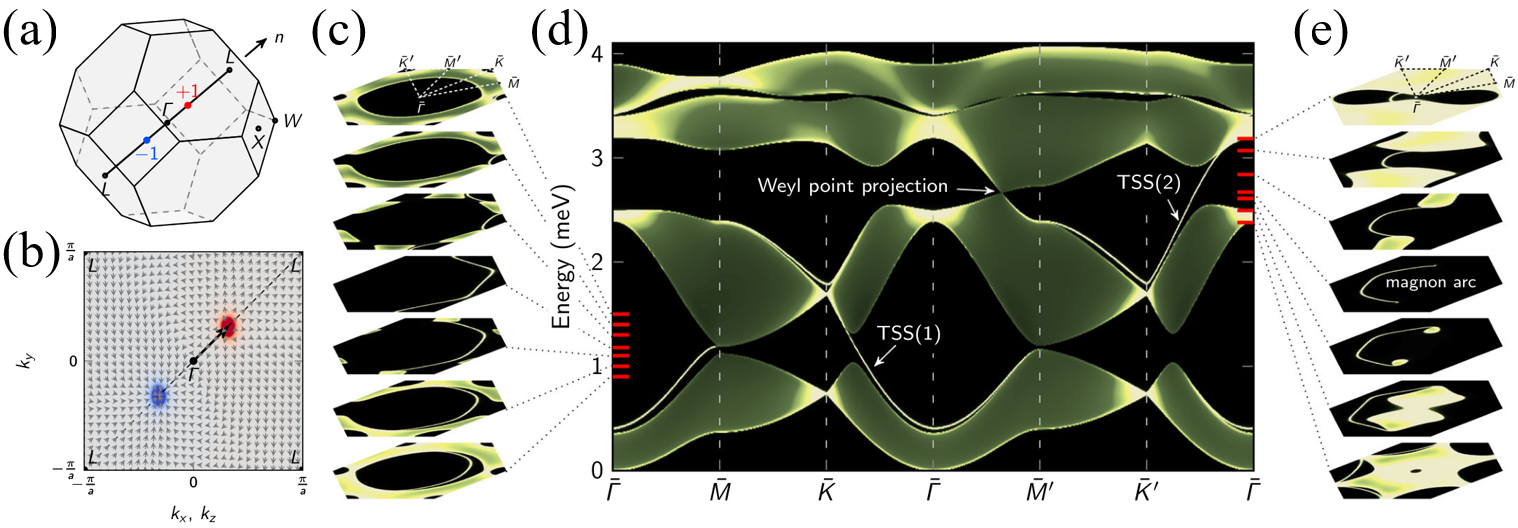}
	\caption{a) Fcc BZ with high-symmetry points, field direction $\boldsymbol n$, and Weyl points (dots). b) Normalized dipole vector field (Berry curvature) of band 2 in the $k_x=k_z$ plane. The color scale depicts the divergence of the vector field (blue: negative; gray: zero; red: positive); the two Weyl points appear in the center of the blue and red spots, respectively. c) Magnon band structures at the (111) surface. The surface spectral density is shown as a color scale (black: zero; white: maximum). c) and e show constant-energy cuts through the entire surface Brillouin zone for energies indicated by red lines in d. d) Spectral density along high-symmetry directions of the surface Brillouin zone. Adapted with permission. \cite{Mook2016} Copyright 2016, American Physical Society.}
	\label{fig:3-6}
\end{figure*}
\subsection{$\mathbb{Z}_2$ Topological Magnon Insulators}
In general, topological phases are characterized by their topological invariants. The topological invariant of the Chern insulator is the (first) Chern number or TKNN invariant \cite{Thouless1982,Kohmoto1985}, whose chiral edge states one-to-one correspond to the value of the Chern number in the integer quantum Hall effect with broken time-reversal symmetry \cite{Klitzing1980,Hatsugai1993}. Nevertheless, it was gradually realized that this bulk-edge correspondence can also happen in systems with unbroken time-reversal symmetry over the last two decades. In 2015, Kane and Mele \cite{Kane2005,Kane2005b} proposed that the intrinsic spin-orbit interaction can open up a band gap at the Dirac points in graphene without a magnetic field, which plays the role of the magnetic flux in Haldane's model with broken time-reversal symmetry \cite{Haldane1988}. Then the system exhibits a quantum spin Hall effect (QSHE) characterized by a pair of spin-helical gapless edge states, which allow electrons with opposite spins to propagate in opposite directions. This Kane-Mele model can be equivalent to two copies of the quantum anomalous Hall effect (QAHE) with opposite spins in Haldane's model, so that the total system still holds the time-reversal symmetry. That is, these helical edge states are robust against weak disorders by time-reversal symmetry. This insulating phase is therefore classified as a kind of symmetry protected topological insulator \cite{Chiu2016}, i.e. the $\mathbb{Z}_2$ topological insulator, whose topological invariant is characterized by the $\mathbb{Z}_2$ topological order \cite{Kane2005b}. 

On the other hand, a few recent theoretical works have reported the realization of $\mathbb{Z}_2$ topological magnon insulators, where helical magnon edge states protected by an (effective) time-reversal symmetry are expected to exist. Nakata et al. \cite{Nakata2017} established the first magnonic counterpart model of $\mathbb{Z}_2$ topological insulators in semiconductors, who extended the notion of symmetry protected topological phases to antiferromagnetic insulators with the magnetic N\'eel order due to the electric field gradient-induced Aharonov-Casher (AC) effect. Under the assumption that the $z$ component of the total spin $S^z$ remains a good quantum number, this conservation law plays the role of the time-reversal symmetry (which is broken by the antiferromagnetic order) and protects a pair of magnonic helical edge states. The dynamics of magnons in a collinear antiferromagnet can be described as the combination of two independent copies of magnons in a ferromagnet \cite{Nakata2017b} for each mode $\sigma=\pm 1$ (see \Figure{fig:3-5}a). Driven by the AC effect induced by an electric field gradient, up and down magnons with the same frequency perform cyclotron motion in opposite directions (see \Figure{fig:3-5}b), which bring about a pair of magnonic helical edge states in the band gap (see \Figure{fig:3-5}c). \Figure{fig:3-5}b shows a magnonic version of QSHE, and the antiferromagnetic system becomes a $\mathbb{Z}_2$ topological magnon insulator characterized by the $\mathbb{Z}_2$ topological invariant. Meanwhile, light can also control the topological phases of magnon through the AC effect induced by a laser electric field \cite{Nakata2019}. Both linearly and circularly polarized lasers can generate magnonic helical edge states, but the difference is that a linearly polarized laser gives the magnon spin Nernst effect and a circularly polarized one shows the magnon thermal Hall effect. 

Subsequently, several models of $\mathbb{Z}_2$ topological magnon insulator and magnonic QSHE have been proposed, such as in a collinear antiferromagnet on a square-octagon lattice \cite{Mook2018} or a honeycomb lattice \cite{Lee2018} with DM interaction, a canted collinear antiferromagnet on a square lattice \cite{Kawano2019}, A-type antiferromagnet on the kagome bilayer system and G-type antiferromagnet on the honeycomb bilayer system \cite{Kondo2019}. In addition, magnonic three-dimensional $\mathbb{Z}_2$ topological phases have been realized in a diamond lattice system having two spins at each site \cite{Kondo2019b} and AA-stacked honeycomb ferromagnets with antiferromagnetic interlayer coupling \cite{LiY2022}. Yet it is worth noting that an open gap is essential to obtain a magnonic $\mathbb{Z}_2$ topological phases with helical edge states in all of these models. 

\begin{figure}[t]
	\centering
	\includegraphics[width=0.48\textwidth]{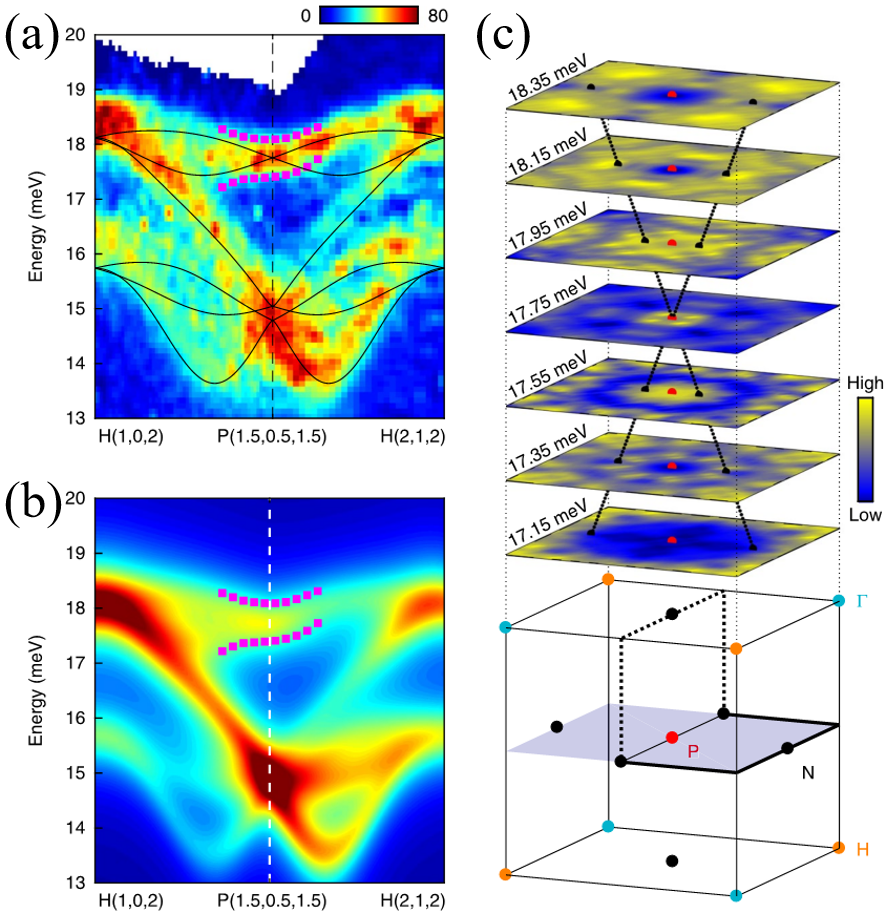}
	\caption{a) and b) Inelastic neutron scattering and calculated $S(\boldsymbol Q,\omega)$, respectively, along an H–P–H momentum trajectory. c) Inelastic neutron scattering intensity distribution in 0.2meV intervals in $\boldsymbol Q$-space planes that connect P with its four neighboring N-points (P–N planes). Adapted with permission. \cite{YaoW2018} Copyright 2018, Springer Nature.}
	\label{fig:3-7}
\end{figure}
\subsection{Topological Magnon Semimetals}
Different from a topological insulator characterized by gapless surface states inside the bulk band gap, the topological semimetal is a new class of quantum materials without the band gaps, which show linear dispersion around nodes \cite{Chiu2016}. For instance, the bands of the Weyl semimetals disperse linearly in momentum space through a Weyl point \cite{XuS2015,LvB2015}, which carry topological charges (a nonzero Chern number) that act as monopoles with a fixed chirality. The Berry curvature becomes singular at these Weyl points, or such a Weyl point can be a source ($+$ chirality) or a sink ($-$ chirality) of the Berry curvature. Weyl points always come in pairs in a Weyl semimetal with the requirement of either the time-reversal symmetry broken or the inversion symmetry broken \cite{WanX2011}. When the time-reversal symmetry is broken but the inversion symmetry is not, each Weyl point at $\boldsymbol k$ with topological charge $q$ has its opposite partner with $-q$ at $-\boldsymbol k$, and the minimal number of Weyl points is two. When the inversion symmetry is broken but the time-reversal symmetry is not, there must exist pairs of Weyl points with opposite topological charges at the same place in momentum space. That is, there are at least four Weyl points. Additionally, when the time-reversal symmetry and inversion symmetry coexist, the bands cross at a two-fold spin degenerate Dirac cone, that the node is a four-fold degeneracy of the Dirac point leading to a Dirac semimetal \cite{Young2012}. In other words, Dirac semimetals can be viewed as Weyl semimetals without symmetries breaking. 

Over the years, the concepts of Dirac or Weyl points have been widely extended to magnon spectrums. The first model of Weyl magnons was constructed in ferromagnetic pyrochlores \cite{Mook2016,SuY2017}, where the (pseudo-spin) time-reversal symmetry is broken by the symmetry-allowed DM interaction. The projections of each pair of magnonic Weyl points onto a surface are connected by magnon arcs due to the topologically protected surface states as shown in \Figure{fig:3-6}. Subsequently, several works proposed Weyl magnons in breathing pyrochlore antiferromagnets \cite{LiF2016,LiF2018,JianS2018,Hwang2020}, stacked honeycomb ferromagnets \cite{SuY2017b,Zyuzin2018,LiB2018}, stacked kagome antiferromagnets \cite{Owerre2018}, rare-earth double perovskites \cite{LiF2017}, and multiferroic ferrimagnet Cu$_2$OSeO$_3$ \cite{ZhangL2020}. Naturally, the time-reversal symmetry is broken due to the magnetic order in magnets. Meanwhile, Kramer’s degeneracy is not applicable for magnons, because magnons are integer bosonic excitations with $\mathcal{T}^2=+1$. These suggest that Weyl points may be generically familiar in magnonic systems. 

On the other hand, Fransson et al. \cite{Fransson2016} show that Dirac magnons are inherent in a two-dimensional magnet on the honeycomb lattice due to the  spatial sublattice symmetry. A ferromagnet exhibits magnonic Dirac points at high-symmetry $K$ and $K'$ points in
the reciprocal space, while magnonic Dirac points transfer to $\Gamma$ point in an antiferromagnet. These Dirac magnons are proven to be robust against magnon-magnon interactions, which tend to only rigidly shift the band structure. Pershoguba et al. \cite{Pershoguba2018}  underlined the role of interacting Dirac magnons, who demonstrate that magnon-magnon interactions give rise to a significant momentum-dependent renormalization of the band structure and strongly momentum-dependent magnon lifetimes. Besides, Dirac magnons have also been proposed in a collinear or noncollinear kagome antiferromagnet \cite{Owerre2017,Okuma2017}. Whereas Weyl magnons lack experimental evidences, Dirac magnons are widely observed by performing inelastic neutron scattering measurements in recent experiments, such as in a two-dimensional van der Waals honeycomb crystal ferromagnet CrX$_3$ (X = I, Br) \cite{ChenL2018,ChenL2021,CaiZ2021,Costa2020} and CrXTe$_3$ (X = Si, Ge) \cite{Zhu2021}, honeycomb-lattice antiferromagnet BaNi$_2$(AsO$_4$)$_2$ \cite{GaoB2021}, stacked honeycomb lattice magnet CoTiO$_3$ \cite{YuanB2020}, and three-dimensional antiferromagnet Cu$_3$TeO$_6$ (see \Figure{fig:3-7}) \cite{YaoW2018,BaoS2018}.  

Overall, the realization of Weyl and Dirac magnon semimetals offers new opportunities for experimental observation of the bosonic topological semimetals and edge states. The Weyl magnon semimetal featured by Weyl points can be detected by inelastic neutron scattering which has been used to probe the Dirac magnons. For magnon arcs and magnonic surface states, it is possible to detect them by using surface-sensitive probe techniques, such as high-resolution electron energy-loss spectroscopy, or spin-polarized scanning tunneling microscopy \cite{Balashov2006}. Besides, as the Weyl magnons will result in the spin Hall and anomalous thermal Hall effects \cite{Zyuzin2018,Owerre2018}, they may be probed by measuring the spin and heat conductances.

\begin{table*}\label{tab1} 
 \caption{Possible materials for realizing topological magnons. (In this table, FM, FiM and AFM represent ferromagnet, ferrimagnet and antiferromagnet.)}
\centering
\tabcolsep=0.024\linewidth
\begin{tabular}{l c c c c}
   \hline
    Material & Crystallographic structure & Magnetism & Remarkable feature \\
    \hline
    Lu$_2$V$_2$O$_7$  & Pyrochlore & FM &  Magnon thermal Hall effect \cite{Onose2010,Ideue2012} \\
      &   &   &  Weyl magnons \cite{Mook2016,SuY2017}  \\
    Cu[1,3-bdc] & Stacked kagome & FM & Magnon thermal Hall effect \cite{Hirschberger2015,Chisnell2015}  \\
    &   &   & Magnon Chern insulator \cite{Chisnell2015,Zhang2013,Mook2014}  \\
    CrXTe$_3$ (X = Si, Ge)  &  Stacked honeycomb  &  FM/AFM & Dirac magnons \cite{Zhu2021}  \\ CrX$_3$ (X = I, Br)  &  Stacked honeycomb  &  FM/AFM & Dirac magnons \cite{ChenL2018,ChenL2021,CaiZ2021,Costa2020}  \\
    &   &   & High-order topological magnon insulator \cite{Mook2021}  \\
    &   &   &  Moir\'e magnons \cite{Ganguli2023,WangH2023}  \\
    $\alpha$-RuCl$_3$ & Kitaev-honeycomb & FM/Spin liquid &  Magnon thermal Hall effect \cite{Cookmeyer2018,McClarty2018,Joshi2018} \\
    &   &   & Magnon Chern insulator \cite{McClarty2018,Joshi2018}  \\
    Cu$_2$OSeO$_3$  &  Pyrochlore & FiM  &  Weyl magnons \cite{ZhangL2020}  \\
    MnPS$_3$ & Stacked honeycomb & AFM &  Magnon Spin Nernst effect \cite{ChengR2016,Zyuzin2016,Shiomi2017} \\
    &   &   & $\mathbb{Z}_2$ topological magnon insulator \cite{Kondo2019,LiY2022}  \\
    &   &   &  Moir\'e magnons \cite{LiY2020}  \\
    BaNi$_2$(AsO$_4$)$_2$  &  Stacked honeycomb  &  AFM & Dirac magnons \cite{GaoB2021}  \\
    CoTiO$_3$  &  Stacked honeycomb  &  AFM & Dirac magnons \cite{YuanB2020}  \\
    Cu$_3$TeO$_6$  &   Centro-symmetric cubic  &  AFM & Dirac magnons \cite{BaoS2018,YaoW2018}  \\
    Eu$_2$Ir$_2$O$_7$&  Pyrochlore & AFM &  Magnon thermal Hall effect \cite{Laurell2017,Hwang2020}  \\
      &   &   &  Weyl magnons \cite{LiF2016,JianS2018}  \\
    CaCu$_3$(OH)$_6$Cl$_2$$\cdot$0.6H$_2$O & Kagome & AFM &  Magnon thermal Hall effect \cite{Doki2018}  \\
    \hline
  \end{tabular}
\end{table*}
\section{Survey of Candidate Materials for Topological Magnons}
From an experimental point of view, the research in topological magnons is still in its infancy and has been limited to a handful of materials. Fortunately, there have been a number of significant achievements in the field, for instance, the observations of the magnon thermal Hall effect and magnonic Dirac points. Recently, Karaki et al. \cite{Karaki2023} presented an efficient symmetry-based approach for searching topological magnons in magnetically ordered crystals. After carrying out a search among 198 compounds with an over 300K transition temperature, 12 magnetic insulators supporting room-temperature topological magnons have been identified. Here, we summarize recent works and give a list of candidate Materials that are currently being intensively investigated for topological magnons in Table 1. However, this list is far from comprehensive and only aims to illustrate the diversity of topological magnons.

In addition to the candidates in real materials mentioned above, we highlight three further artificial magnetic candidate materials in which topological magnon phases have been theoretically proposed and experimentally realized. The first such artificial material is the magnonic crystals \cite{Chumak2015,Krawczyk2014}. Owing to its periodic structure, the spin-wave volume-mode spectrum of the magnetostatic spin wave with the longer wavelength due to the long-range dipolar interaction forms allowed frequency bands of spin-wave states (bulk bands) and forbidden-frequency bands (band gaps). A wide variety of parameters, such as the width, thickness, and saturation magnetization of the sample, can be used to tune the spin-wave band structure. Thus is a concept of band engineering in the magnonic system. As mentioned in Section 3.1, a magnonic topological insulator with topologically protected chiral edge states in a magnonic crystal has been proposed \cite{Shindou2013,Shindou2013b,Shindou2014}. Xu et al. \cite{XuB2016} proposed magnonic analogs of integer quantum Hall states in a two-dimensional spin-ice model with disorders, where the magnon bands show a direct transition from an integer quantum magnon Hall regime to a conventional magnon localized regime. Iacocca et al. \cite{Iacocca2017} calculate the spin-wave band structure for square artificial spin ices composed of geometrically placed magnetic nanoislands coupled through dipolar interactions, where an interfacial DM interaction was taken into account by an adjacent heavy-metal layer. The topologically magnonic edge states due to the interfacial DM interaction can be easily tuned by spin configurations in magnetic nanoislands. Hu et al. \cite{Hu2022} realized topological magnonic surface states in antiparallelly aligned magnetic multilayers, who demonstrated that the bulk bands with nonzero Chern numbers and magnonic surface states in the band gaps carrying chiral spin currents are generated by the long-range chiral interlayer dipolar interaction. The surface states are highly localized and can be easily switched between nontrivial and trivial phases by applying an external magnetic field. Most recently, Feilhauer et al. \cite{Feilhauer2023} numerically demonstrated unidirectional, topologically protected edge states in a magnonic crystal composed of dipolar coupled Permalloy triangles. The system undergoes a couple of topological phase transitions by tuning the strength of the perpendicular magnetic field, which gives rise to the change of direction of the topological edge state. 

The second artificial material is a magnet with topological spin textures. In Section 3.3, we mentioned that a second-order topological magnon insulator characterized by magnonic corner states has been predicted in a two-dimensional antiskyrmion crystal \cite{Hirosawa2020}. Additionally, the topological magnon and its thermal Hall effect have been demonstrated in a ferromagnetic \cite{Molina2016,Diaz2020}, antiferromagnetic \cite{Diaz2019}, and ferrimagnetic \cite{Kim2019b} skyrmion crystal. In this case, the fictitious magnetic fields due to the equilibrium magnetic texture act as the effective spin-orbit coupling leading to the topologically protected magnonic edge state \cite{Hoogdalem2013}. So far, only one experimental work investigated the topological magnon band structure in a lattice of skyrmion tubes in manganese silicide by performing the polarized inelastic neutron scattering \cite{Weber2022}.

The last artificial material is moir\'e superlattices comprising twisted bilayers of van der Waals magnets, for instance, chromium triiodide CrX$_3$ (X=I, Br). Recently, signatures of magnetic ground states in twisted (double) bilayer CrI$_3$ have been identified with micro-Raman spectroscopy measurements \cite{XieH2022}. Meanwhile, a magnetic skyrmion bubble with non-conserved helicity was predicted in twisted bilayer CrI$_3$ \cite{YangB2023}. Pioneering theoretical efforts have predicted that topological magnons can be realized by twisting bilayer magnets \cite{WangC2020,LiY2020,Ghader2021,HuaC2023}. So far, several experimental works have investigated the magnon band structure identifying the properties of moir\'e magnons \cite{ChenJ2022,Ganguli2023,WangH2023}. And one of them observed the magnonic edge modes at an optimal twist angle and with a selective excitation frequency \cite{WangH2023}.  

\section{Summary and Outlook}
In summary, we have provided an overview of the most recent research on the topological phases of magnons in magnonic systems, including Chern insulators, high-order topological insulators, $\mathbb{Z}_2$ topological insulators, and topological semimetals. Throughout this review, we devoted ourselves to building a bridge between topology and magnonics utilizing the existing results of topological physics in electronic systems. We first introduce some basic notions and necessary theoretical fundamentals, which are essential for readers to understand this topic. Then to systematically summarize the previous research and generalize the main focus, we have highlighted several important achievements in the field of topological magnonics mostly in the past decade. 

Although a vast number of model studies and theoretical predictions have been made over the past decade, few experimental realizations for topological magnons have been achieved so far. Because experimental work on topological magnons is still at a relatively early stage, only a handful of materials have been verified for topological magnons. Therefore, we summarized candidate materials including some artificial structures for readers to get a better understanding of topological magnons. As the Weyl magnons and high-order topological magnons have not been observed, another pressing problem is to develop more advanced technology to detect the topological surface states of magnons. Overall, the realization of topological magnons will offer new opportunities for experimental observation of the topological phases and edge states. We believe that rapid progress in the field of topological magnonics will greatly deepen the understanding of topological physics in condensed matter.

\section*{Acknowledgements} 
F.Z. and J.K. acknowledge the supports from the Postdoctoral International Exchange Program of China (Grant No.YJ20220302), the National Natural Science Foundation of China (Grant No.12074276), the Double First-Class Initiative Fund of ShanghaiTech University, and the start-up grant of ShanghaiTech University. A.M. acknowledges support from the Excellence Initiative of Aix-Marseille Universit\'e---A*Midex, a French "Investissements d'Avenir" program. Z.X.C. thanks Australia Research Council for support (DP190100150). 

\section*{Conflict of Interest} 
The authors declare no conflict of interest.

\section*{Keywords} 
Topological phase, magnon, Hall effect, edge state, topological insulator, topological semimetal.

\end{document}